\documentstyle[epsfig]{mn}

\newif\ifAMStwofonts

\ifoldfss
  \ifCUPmtlplainloaded \else
    \NewTextAlphabet{textbfit} {cmbxti10} {}
    \NewTextAlphabet{textbfss} {cmssbx10} {}
    \NewMathAlphabet{mathbfit} {cmbxti10} {} 
    \NewMathAlphabet{mathbfss} {cmssbx10} {} 
  \fi
  \ifAMStwofonts
    \ifCUPmtlplainloaded \else
      \NewSymbolFont{upmath} {eurm10}
      \NewSymbolFont{AMSa} {msam10}
      \NewMathSymbol{\upi}     {0}{upmath}{19}
      \NewMathSymbol{\umu}     {0}{upmath}{16}
      \NewMathSymbol{\upartial}{0}{upmath}{40}
      \NewMathSymbol{\leqslant}{3}{AMSa}{36}
      \NewMathSymbol{\geqslant}{3}{AMSa}{3E}

       \let\le=\leqslant
       \let\ge=\geqslant
    \fi
  \fi
\fi 

\ifnfssone
  \newmathalphabet{\mathit}
  \addtoversion{normal}{\mathit}{cmr}{m}{it}
  \addtoversion{bold}{\mathit}{cmr}{bx}{it}
  \newmathalphabet{\mathbfit} 
  \addtoversion{normal}{\mathbfit}{cmr}{bx}{it}
  \addtoversion{bold}{\mathbfit}{cmr}{bx}{it}
  \newmathalphabet{\mathbfss} 
  \addtoversion{normal}{\mathbfss}{cmss}{bx}{n}
  \addtoversion{bold}{\mathbfss}{cmss}{bx}{n}
  \ifAMStwofonts
    \ifCUPmtlplainloaded \else
      \UseAMStwoboldmath
      \makeatletter
      \new@mathgroup\upmath@group
      \define@mathgroup\mv@normal\upmath@group{eur}{m}{n}
      \define@mathgroup\mv@bold\upmath@group{eur}{b}{n}
      \edef\UPM{\hexnumber\upmath@group}
      \new@mathgroup\amsa@group
      \define@mathgroup\mv@normal\amsa@group{msa}{m}{n}
      \define@mathgroup\mv@bold\amsa@group{msa}{m}{n}
      \edef\AMSa{\hexnumber\amsa@group}
      \makeatother
      \mathchardef\upi="0\UPM19
      \mathchardef\umu="0\UPM16
      \mathchardef\upartial="0\UPM40
      \mathchardef\leqslant="3\AMSa36
      \mathchardef\geqslant="3\AMSa3E

       \let\le=\leqslant
       \let\ge=\geqslant
    \fi
  \fi
\fi 

\ifnfsstwo
  \DeclareMathAlphabet{\mathbfit}{OT1}{cmr}{bx}{it}
  \SetMathAlphabet\mathbfit{bold}{OT1}{cmr}{bx}{it}
  \DeclareMathAlphabet{\mathbfss}{OT1}{cmss}{bx}{n}
  \SetMathAlphabet\mathbfss{bold}{OT1}{cmss}{bx}{n}
  \ifAMStwofonts
    \ifCUPmtlplainloaded \else
      \DeclareSymbolFont{UPM}{U}{eur}{m}{n}
      \SetSymbolFont{UPM}{bold}{U}{eur}{b}{n}
      \DeclareSymbolFont{AMSa}{U}{msa}{m}{n}
      \DeclareMathSymbol{\upi}{0}{UPM}{"19}
      \DeclareMathSymbol{\umu}{0}{UPM}{"16}
      \DeclareMathSymbol{\upartial}{0}{UPM}{"40}
      \DeclareMathSymbol{\leqslant}{3}{AMSa}{"36}
      \DeclareMathSymbol{\geqslant}{3}{AMSa}{"3E}

       \let\le=\leqslant
       \let\ge=\geqslant
    \fi
  \fi
\fi 

\ifCUPmtlplainloaded \else
  \ifAMStwofonts \else 
    \def\upi{\pi}
    \def\umu{\mu}
    \def\upartial{\partial}
  \fi
\fi

\title{Cosmic  star formation: constraints on the galaxy formation models}
\author[F. Calura, F. Matteucci, N. Menci]
       {F. Calura$^{1}$\thanks{E-mail: fcalura@ts.astro.it}, 
        F. Matteucci$^{1}$, N. Menci$^{2}$\\
        (1) Dipartimento di Astronomia-Universit\'a di Trieste, Via G.
B. Tiepolo
	11, 34131 Trieste, Italy\\
	(2) INAF, Osservatorio Astronomico di Roma, via Frascati 33, I-00040 Monteporzio, Italy\\
	 }
	
\date{Accepted ---- .
      Received ---- ;
      in original form ----}

\pagerange{\pageref{firstpage}--\pageref{lastpage}}
\pubyear{2004}

\begin{document}

\maketitle

\label{firstpage}
\begin{abstract}
We study the evolution of the cosmic star formation in the universe by computing the 
luminosity density (in the UV, B, J, and K bands) and the stellar mass density of 
galaxies in two reference  models of  galaxy evolution: the  pure-luminosity evolution  (PLE)
model developed by Calura \& Matteucci (2003) and the semi-analytical  model
(SAM)  of hierarchical  galaxy formation  by Menci  et al.  (2002). The  former
includes  a  detailed  description  of the  chemical  evolution  of  galaxies of
different morphological types; it does not include any number evolution of 
galaxies whose number density is normalized to the observed local value. 
On the other hand, the  SAM includes a strong density evolution following the 
formation and the merging histories of the DM haloes hosting the galaxies, as 
predicted by the hierarchical clustering scenario, but it does not contain morphological 
classification nor chemical evolution.  
Our results suggest that at low-intermediate 
redshifts ($z< 1.5$) both models are consistent with the available data 
on the luminosity density of galaxies in all the considered bands.  
At high redshift the luminosity densities predicted in the PLE model  
show a peak due to the formation of ellipticals, whereas in the hierarchical picture 
a gradual decrease of the star formation and of the luminosity densities is predicted 
for $z> 2.5$. At such redshifts the PLE predictions tend to overestimate  
the present data in the B band whereas the SAM tends to underestimate the 
observed UV luminosity density. As for the stellar mass density, 
the PLE  picture predicts that nearly   $50 \%$ and $85  \%$ of the
present  stellar  mass  are in place  at  $z\sim  4$  and $z\sim1$,  respectively.
According to the hierarchical SAM, $50 \%$  and  $60 \%$ of the present  stellar
mass are completed at $z \sim  1.2$ and $z=1$, respectively. Both  
predictions  fit  the  observed  stellar  mass density 
evolution up to $z=1$. At $z>1$, the PLE and SAM models tend to overestimate and
underestimate the  observed values,  respectively. 
We discuss the origin of the 
similarities and of the discrepancies between the two models, and 
the role of observational uncertainties (such as dust extinction) in 
comparing models with observations.

\end{abstract} 


\section{Introduction}
In the past few years a great deal of work appeared 
on the subject of galaxy formation and evolution. 
With the  word  ''formation'' usually
one means the assembly of the bulk of the material (say $> 50 \%$) of
the luminous part  of a galaxy, namely the stars and the gas, within a sphere of
radius of $\sim$30 kpc (Peebles 2003).  A reliable picture of
galaxy formation must be able to reproduce, at the same time, all (or
as much as possible) of the available constraints, including colors and chemical abundances. 
Currently, the most intriguing debate on galaxy evolution concerns how the
formation of ellipticals and bulges occurred in the universe.  In fact, the two
main competing scenarios of galaxy evolution propose rather different
conditions for the formation of spheroids.  In the first scenario,
ellipticals and bulges formed at high redshift (e.g. $z > 2-3$) as the
result of a violent burst of  star formation following a ``monolithic
collapse'' (MC) of a gas cloud. 
After the main burst of star formation, the galaxy lost the residual gas by means of 
a galactic wind and it evolved passively since then 
(Larson 1974, van Albada 1982, Sandage 1986, Matteucci \& Tornamb\'e 1987, 
Arimoto \& Yoshii 1987, Matteucci 1994). 
The monolithic collapse view, or better, the idea that spheroids formed quickly and at high redshift, 
 is supported by a large set of observational evidences. Among them, of particular importance are  
the thinness of the Fundamental Plane (Djorgovski \& Davis 1987,  Renzini 
\& Ciotti 1993, Bernardi et al. 1998, Kochanek et al. 2000, van Dokkum et
al. 2001, Rusin et al. 2003, van Dokkum \& Ellis 2003),  
the overabundance of Mg relative to Fe observed in the stars as well as the increase of the [Mg/Fe] ratio with galaxy 
luminosity (Pipino \& Matteucci 2004 and references therein), 
the tightness of the color-central velocity 
dispersion and color-magnitude relation (Bower, Lucey \& Ellis 1992, Kodama et al. 1999) 
observed for both cluster and field spheroids at high and low redshift, 
as well as the constancy of the number density of both spheroids and large discs observed up to $z\sim 1$ 
(Im et al. 1996, Lilly et al. 1998, Schade et al. 1999, Im et al. 2002).\\ 
On the other hand, the hierarchical clustering (HC) picture is based on the Press \& Schechter 
(1974) structure 
formation theory, which has been developed mainly to study the behaviour of the dark matter.  
According to this theory, 
in a $\Lambda$-Cold dark Matter ($\Lambda$CDM)-dominated   
universe, small DM halos are the first to collapse, then interact and merge 
to form larger halos. 
The most uncertain assumption in the HC scenario concerns the behaviour of the baryonic matter, 
which is assumed to follow the DM in all the interaction and merging processes. 
In this framework, massive spheroids are formed from 
several merging episodes among gas-rich galaxies,  such as discs,
occurring throughout the whole Hubble time.  These mergers produce
moderate star formation rates (SFRs), with massive galaxies reaching their
final masses at more recent epochs than less massive ones 
($z \le 1.5$, White \& Rees 1978, 
Kauffmann, White \& Guiderdoni 1993, Baugh et al. 1998, Cole et
al. 2000, Somerville et al. 2001, Menci et al. 2002). 
The observational evidence in favor of the hierarchical galaxy formation is  
the apparent paucity of giant galaxies at high redshift ($z\sim 1$) as claimed by some authors (Barger et al. 1999, Kauffmann,
Charlot \& White 1996, Zepf 1997), the blue colors of some spheroids at low redshift, possibly ascribed to residual star 
formation 
activity induced by mergers (Franceschini et al. 1998, Menanteau et al. 1999),  
as well as the observations showing evidence for mergers in distant field and cluster galaxies (Bundy et al. 2004, 
van Dokkum et al. 2000) 
and the increase of the measured merging rate with redshift (Patton et al. 1997, Le F\'evre et al. 2000, 
Conselice et al. 2003).\\
Recently, Calura \& Matteucci (2003, hereinafter CM03) have developed a series of detailed chemical and 
spectro-photometric models for elliptical, spiral and irregular galaxies, used to 
study the evolution of the luminous matter 
in the universe and the contributions that galaxies of different morphological types 
bring to the 
overall cosmic star formation. It is worth noting that all these models reproduce the chemical 
abundances and abundance patterns in the aforementioned galaxies.

In their scenario of pure-luminosity evolution (PLE), 
only the galaxy luminosities evolve, whereas the number densities are assumed to be 
constant and equal to the values indicated by the local B-band 
luminosity function (LF), as observed by Marzke et al. (1998). 
In this paper, we compare the cosmic star formation history as predicted 
by the PLE model of CM03 with the predictions of the hierarchical semianalytic model (SAM)  
developed by Menci et al. (2002).  
We want to stress that the PLE model and the SAM do not represent the only alternatives to study galaxy evolution. 
For instance, several groups study  the evolution of the cosmic star formation by means of large-scale hydrodynamical simulations
(e.g. Sringel \& Hernquist 2003, Nagamine et al. 2004), which are generally based on the $\Lambda$CDM cosmological model. 
However, representing the PLE and Menci SAM considered in this work two rather opposite scenarios and 
providing rather extreme predictions, they may be helpful to constrain 
the parameter space also for other galaxy formation models. 
Furthermore, we want to stress that not necessarily the two scenarios are in contradiction,  
since the HC was devised for the DM whereas the PLE for the baryonic matter. 
As some observational evidence seems to indicate, it is in fact  
possible that, although DM halo formation is hierarchical, 
the baryonic matter evolved in an anti-hierarchical fashion, in the sense 
that larger galaxies are older than small ones (Matteucci 1994, Pipino \& Matteucci 2004). 
By comparing the model predictions with a large set of observational data, we aim at  
inferring whether the two main competing scenarios can be disentangled on the basis of the current observations. 
The novelty with respect to the paper by CM03 is the incorporation of dust extinction in the 
PLE model,  
with important consequences on the predicted behaviour of the luminosity of galaxies at short wavelengths, 
i.e. in the UV and B photometric bands. 
This paper is organized as follows: in sections 2 and 3, we describe the pure-luminosity evolution model as developed by 
CM03  and the SAM by Menci et al. (2002), respectively. 
In section 4 we present our results, and in section 5 we draw 
the conclusions. Unless otherwise stated, throughout the paper we use a $\Lambda$CDM cosmological 
model characterized by $\Omega_{0}=0.3$, $\Omega_{\Lambda}=0.7$ and $h=0.65$.

\section{The CM03 pure-luminosity evolution model}
The PLE models developed by CM03 consist of chemical evolution models for galaxies of different morphological 
types (ellipticals, spirals, irregulars), used to calculate metal abundances and star formation rates (SFRs), 
and by a spectro-photometric code used to calculate galaxy spectra, colors and magnitudes 
by taking into account the chemical evolution.  
Detailed descriptions of the chemical evolution models for galaxies of different morphological types 
can be found in Matteucci
\& Tornamb\'{e} (1987) and Matteucci (1994) for elliptical galaxies, 
Chiappini et al. (1997, 2001) for the spirals and 
Bradamante et al. (1998) for irregular galaxies. 
We assume that the category of galactic bulges is naturally included in the one of elliptical galaxies. 
Our assumption is supported by the similar features characterizing bulges and ellipticals: 
for instance, 
both are dominated by old stellar populations and respect the same fundamental plane (Binney \&
Merrifield 1998, Renzini 1999). 
This indicates that they are likely to have a common origin, i.e. both are likely to have
formed on very short timescales and a long time ago, and we will refer to both ellipticals and bulges as to the ``spheroids''.\\
In our picture, spheroids form as a result of the rapid collapse of a homogeneous sphere of
primordial gas where star formation is taking place at the same time as the collapse proceeds. 
Star formation is assumed to halt as the energy of the ISM, heated by stellar winds and SN explosions,
balances the binding energy of the gas. At this time a galactic wind occurs, sweeping away almost all of  
the residual gas. By means of the galactic wind, ellipticals enrich the inter-galactic medium (IGM) with metals.\\ 
For spiral galaxies, the adopted model is calibrated in order to reproduce a large set of observational 
constraints for the Milky Way galaxy (Chiappini et al. 2001). The Galactic disc is approximated by several independent rings, 
2 kpc wide, without exchange of matter between them. In our picture, 
spiral galaxies are assumed to form as a result of two main infall episodes.  
During the first episode, the halo and the thick disc are formed.
During the second episode, a slower infall
of external gas forms the thin disc with the gas accumulating faster in the inner than in the outer
region ("inside-out" scenario, Matteucci \& Fran\c cois 1989). The process of disc formation is much longer 
than the halo 
and bulge formation, with time scales varying from $\sim2$ Gyr in the inner disc to $\sim8$ Gyr in the solar region
and up to $10-15$ Gyr in the outer disc.\\
In this case, at variance with Chiappini et al. (2001) CM03 assume a Salpeter (1955) IMF, 
instead of the Scalo (1986) IMF. 
This choice is motivated by the fact that a Scalo or a Salpeter IMF in spirals produce very similar results 
in the study of the luminosity density evolution, and also by the fact that we aim to test 
the hypothesis of a universal IMF (see also Calura \& Matteucci 2004).  
Another difference between the Chiappini et al. (2001) model and ours concerns the elimination 
of the star formation threshold, motivated by the fact that its effects 
are appreciable only on 
small scales, i.e. in the chemical evolution of the solar vicinity and of small galactic regions, whereas 
our aim is to study star formation in galactic discs on global scales. 
Finally, irregular dwarf galaxies are assumed to assemble from continuous 
infall of gas  
of primordial chemical composition, until masses in the range $\sim 10^{8} - 6 \times 10^{9}M_{\odot}$ are accumulated, 
and to produce stars at a lower rate than spirals.\\
Let $G_{i}$ be the fractional mass of the element $i$ in the gas
within a galaxy, its temporal evolution is described by the basic equation:\\
\begin{equation}
\dot{G_{i}}=-\psi(t)X_{i}(t) + R_{i}(t) + (\dot{G_{i}})_{inf} - 
(\dot{G_{i}})_{out}\\
\end{equation} 
where $G_{i}(t)=\sigma_{g}(t)X_{i}(t)/\sigma_{tot}$ is the gas mass in the form of an
element $i$ normalized to a total initial mass $M_{tot}$. The quantity $X_{i}(t)=
G_{i}(t)/G(t)$ represents the abundance in mass of an element $i$, with
the summation over all elements in the gas mixture being equal to unity.
The quantity $G(t)= \sigma_{g}(t)/\sigma_{tot}$ is the fractional mass of gas
present in the galaxy at time $t$.
$\psi(t)$ is the instantaneous star formation rate (SFR), namely the fractional amount
of gas turning into stars per unit time; $R_{i}(t)$ represents the returned
fraction of matter in the form of an element $i$ that the stars eject into the ISM through stellar winds and 
SN explosions; this term contains all the prescriptions regarding the stellar yields and
the SN progenitor models.
The two terms 
$(\dot{G_{i}})_{inf}$ and $(\dot{G_{i}})_{out}$ account for the infalling
external gas from the IGM and for the outflow, occurring
by means of SN driven galactic winds, respectively. 
The main feature characterizing a particular morphological galactic type is
represented by the prescription adopted for the star formation history.\\ 
In the case of elliptical and irregular  
galaxies the SFR $\psi(t)$ (in $Gyr^{-1}$) has a simple form and is given by:

\begin{equation}
\psi(t) = \nu G(t) 
\end{equation}

The quantity $\nu$ is the efficiency of star formation, namely the inverse of
the typical time scale for star formation and for ellipticals and bulges is assumed to 
be $\sim 10-15$ Gyr$^{-1}$ (Matteucci 1994). In the case of spheroids, $\nu$ is
assumed to drop to zero at the onset of a galactic wind, which develops as the
thermal energy of the gas heated by supernova explosions exceeds the binding
energy of the gas (Arimoto \& Yoshii 1987, Matteucci \& Tornamb\'{e} 1987). 
This quantity is strongly
influenced by assumptions concerning the presence and distribution of dark
matter (Matteucci 1992); 
for the model adopted here a diffuse 
($R_e/R_d$=0.1, where
$R_e$ is the effective radius of the galaxy and $R_d$ is the radius 
of the dark matter core) but 
massive ($M_{dark}/M_{Lum}=10$) dark halo has 
been assumed.\\
In the case of irregular galaxies we have assumed a continuous star formation rate always expressed as in (2), but
characterized by an efficiency lower than the one adopted for ellipticals, i.e. $\nu =0.01$ Gyr$^{-1}$\\
In the case of spiral galaxies, the SFR expression is:\\
\begin{equation}\label{sfrspi}
\psi(r,t)=\nu \sigma_{tot}^{k_{1}}(r,t) \sigma_{g}^{k_{2}}(r,t)
\end{equation}
where $k_{1}=0.5$ and $k_{2}=1.5$ (see Matteucci \& Fran\c cois 1989, Chiappini et al. 1997). 
For massive stars (M $> 8 M_{\odot}$) 
we adopt nucleosynthesis 
prescriptions by Nomoto et al. (1997a), the yields by van den Hoeck \& Groenewegen (1997) for
low and intermediate mass stars ($0.8 \le M/M_{\odot} \le 8$) and those of 
Nomoto et al. (1997b) for type I a SNe.\\ 
For all galaxies, we assume a Salpeter IMF, expressed by the formula:
\begin{equation}
\phi(m) = \phi_{0} m^{-(1+x)} 
\end{equation}
with $x=1.35$, being the mass range $0.1 \le m/m_{\odot} \le 100$. 
To calculate galaxy colors and magnitudes, we use the photometric code by Bruzual \& Charlot (2003, hereinafter BC). 
However, we have implemented the BC code by taking into account the 
evolution of metallicity in galaxies (Calura 2004).   
Dust extinction is also properly taken into account.  
The chosen geometrical dust distribution plays an important role in the modelling of dust attenuation in galaxies:  
usually, the ``screen'' and ``slab'' dust distributions represent the two most extreme cases.  
In the screen model, the dust is distributed along 
the line of sight of the stars, whereas in the slab model the dust has the same distibution as stars. 
The main difference between the screen and slab dust distributions is the expression of the attenuation factor, 
which in the former case is given by:
\begin{equation}
a_{screen} =  \exp{(-\tau(\lambda))}
\end{equation}
whereas in the latter case it is given by:
\begin{equation}
a_{slab} = [1- \exp{(-\tau(\lambda))}]/\tau(\lambda)
\end{equation}
(Totani \& Yoshii 2000), where $\tau(\lambda)$ is the optical depth of the dust.
In this case, we adopt the ``screen'' geometric distribution which, according to UV and optical 
observations of local starburst galaxies, is to be considered favored over the ``slab'' model 
(Calzetti et al. 1994). 
The absorbed flux $I_{a}(\lambda)$ of a stellar population behind a screen of dust is given by:
\begin{equation}
I_{a}(\lambda)=I_{I}(\lambda) \, \exp{(-\tau(\lambda))}
\end{equation}
(Calzetti 2001), where $I_{I}(\lambda)$ represents the intrinsic, unobscured flux at the wavelength 
$\lambda$.\\
We assume that the optical depth is proportional to the column density $N(g)$ and to the metallicity 
$Z$ of the gas, according to:
\begin{equation}
\tau(\lambda)= C \, k(\lambda) \, N(g) \, Z  
\end{equation}
where $k(\lambda)$ is the extinction curve. 
For spiral galaxies, we adopt the extinction curve derived by Seaton (1979) for the Milky Way (MW) galaxy.  
Such a choice is motivated by the fact that we assume that, as far as the chemical and photometric features are concerned, 
the Milky Way Galaxy represents an average spiral.  
Local starburst galaxies are generally characterized by extinction curves slightly different from the ones of the 
MW (Calzetti 1997, 2001) and are better modelled by the expression found by Calzetti (1997). 
We assume that in the starbursts occurring in elliptical and irregular galaxies the dust follows an attenuation law 
similar to the one estimated by Calzetti (1997) for local starbursts. 
The constant C in equation (6) is chosen in order to reproduce the Milky Way average V-band extinction 
of $A_{V}=0.17$ (Schlegel, Finkbeiner \& Davis 1998).\\
The galaxy densities of the various morphological types are normalized 
according to the local B-band luminosity function observed by Marzke et al. (1998). 
A scenario of pure 
luminosity evolution has been assumed, namely that galaxies evolve 
only in luminosity and not in number. 
This is equivalent to assume that the effects of galaxy interactions and mergers are 
negligible at any redshift. 
Such a picture can account for many observables, such as the evolution 
of the galaxy luminosity density in various bands and the cosmic supernova rates (CM03). 
At redshift larger than zero the absolute magnitudes are calculated according to:
\begin{equation}
M_{B}(z)=M_{B}(z=0)+2.5log(\frac{\int E_{\lambda/1+z}(z)R_{B}(\lambda)d\lambda}{\int
E_{\lambda/1+z}(0)R_{B}(\lambda)
d\lambda})
\end{equation}
where $M_{B}(z=0)$ and $M_{B}(z)$ are the absolute blue magnitudes at redshift 0 and z, respectively, 
$E_{\lambda}(z)\, d\lambda$ is the energy per unit time radiated at the rest-frame wavelength $\lambda$
by the galaxy at redshift $z$, 
and $R_{B}(\lambda)$ is the response function of the rest-frame B band. 
The second term on the right side of equation 7 represents the evolutionary correction (EC), i.e. the
difference in absolute 
magnitude measured in the rest frame of the galaxy at the wavelength of emission (Poggianti 1997).\\
For the LF, we assume a Schechter (1976) form, given by:
\begin{equation}
 \Phi(M) \, dM=0.4 \, ln(10) \, 
 \Phi^{\ast}\, e^{-X} \, X^{\alpha + 1} \, dM 
\end{equation}
where $X=L/L^{\ast}=10^{0.4(M-M^{*})}$. $M^{*}$ ($L^{*}$) is the characteristic magnitude (luminosity) and 
is a function of redshift, whereas  
$\Phi^{\ast}$ and $\alpha$ are the normalization and the faint-end slope, respectively, 
 and are assumed to be constant.\\ 
In bands other than B we assume that the LF shape is the same as in the B band and we 
calculate the LF in the given band (X) transforming the absolute magnitudes according to the 
rest-frame galaxy colors 
as predicted by the spectrophotometric model:\\
\begin{equation}
M_{X}=M_{B}+(X-B)_{rf}
\end{equation}
The LD per unit frequency in a given band (centered at the wavelength $\lambda$) and for the $k-$morphological type is:
\begin{equation}
 \rho_{\lambda,k}= \int{\Phi_{k}(L_{\lambda})\, (L_{\lambda}/L^{\ast}_{\lambda})\, dL_{\lambda}} 
\end{equation}
The total LD is 
given by the sum of the single contributions of spheroids, spirals and irregulars.\\
The stellar mass densities for galaxies of the $k-th$ morphological type 
are $\rho_{*,k}$ and are calculated as: 
\begin{equation} 
\rho_{*,k}=\rho_{B,k} \cdot (M_{*}/L)_{B,k} \\
\end{equation} 
where $\rho_{B,k}$ is the predicted B luminosity density, whereas $(M_{*}/L)_{B,k}$ is the predicted stellar mass to light 
ratio for the $k-$th galactic morphological type. All the galaxies are 
assumed to start forming stars at the same redshift $z_{f}=5$.

\section{The SAM model}
In semianalytical models the galaxy mass distribution is derived from the merging histories of 
the host DM haloes, under the assumption that the galaxies contained in each 
halo coalesce into a central dominant galaxy if their dynamical friction 
timescale is shorter than the halo survival time. The surviving galaxies 
(commonly referred to as satellite galaxies) retain their identity and continue 
to orbit within the halo. The histories of the DM condensations rely on a 
well established framework (the extended Press \& Schechter theory, EPST, see 
Bower 1991; Bond et al. 1991; Lacey \& Cole 1993). However, the recipe 
concerning the 
galaxy fate inside the DM haloes is guided by {\it a posteriori} consistency 
with the outputs of high-resolution N-body simulations. 
The SAM includes the main dynamical processes taking place inside the host DM halos, 
namely dynamical friction and binary aggregations of satellite galaxies. The evolution of the 
galaxy mass distribution is calculated by solving numerically a set of evolutionary equations 
(Poli et al. 1999).\\
The link between stellar evolution and the dynamics follows a procedure widely used in 
semianalythic models. The baryonic content $(\Omega_b/\Omega_m)\,m$ of the galaxy is divided into (1) a 
hot phase  with mass $m_h$ at the virial temperature $T=(1/2)\,\mu\,m_H\,v^2/k$ 
($m_H$ is the proton mass and $\mu$ is the mean molecular weight), 
(2) into a cold phase 
with mass $m_c$ able to radiatively cool within the galaxy survival time, and 
the stars (3) (with total mass $m_*$) forming from the cold phase on a 
time scale 
$\tau_*$. Initially, all baryons are assigned to the hot phase.\\ 
Also in this case, we compute galaxy spectra and luminosities by means of the spectrophotometric code 
developed by Bruzual \& Charlot (2003). 
The integrated stellar emission 
$S_{\lambda}(v,t)$ at the wavelength $\lambda$ for a galaxy of circular velocity $v$ at the time 
$t$ is computed by convolving with 
the spectral energy distribution $\phi_{\lambda}$ obtained from population 
synthesis models: 
\begin{equation}\label{sed}
S_{\lambda}(v,t) = \int_0^t\,dt'\,\phi_{\lambda}(t-t')\,\dot m_*(v,t')~. 
\end{equation}
$\phi_{\lambda}$ is taken from Bruzual \& Charlot (2003), with a Salpeter IMF. 
The metallicity is calculated by assuming a constant effective yield. 
The average galaxy 
metallicity varies between $Z \sim 0.003$ and $Z \sim 0.01$, in agreement with results of 
other SAMs (e.g., Cole et al. 2000). To calculate galactic spectra, we use simple stellar populations (SSPs) at fixed 
metallicity $Z=0.004$. The use of the SSPs at $Z=0.008$ would produce very small 
variations in our results, certainly of negligible entity with respect to the observational errors.\\
The dust extinction affecting the above luminosities is 
computed assuming the dust optical depth to be proportional to the metallicity 
$Z_{cold}$ of the cold phase and to the disk surface density, so that for the $V$-
band $\tau_{V}\propto m_c\,Z_{cold}/\pi\,r_d^2$. The proportionality constant is 
taken as a free parameter chosen to fit the bright end of the local LF.  
This fact yields, for the proportionality constant, the 
value 
$3.5\,M_{\odot}^{-1}\,{\rm pc}^2$ with the stellar yield producing a 
solar metallicity for a $v=220$ km/s galaxy. 
Physically, this recipe for computing dust extinction is identical to the one used for the PLE model (eq. 6). 
To compute the extinction in the 
other bands, we use the extinction law of Calzetti (1997). 

\begin{figure*}
\centering
\vspace{0.001cm}
\epsfig{file=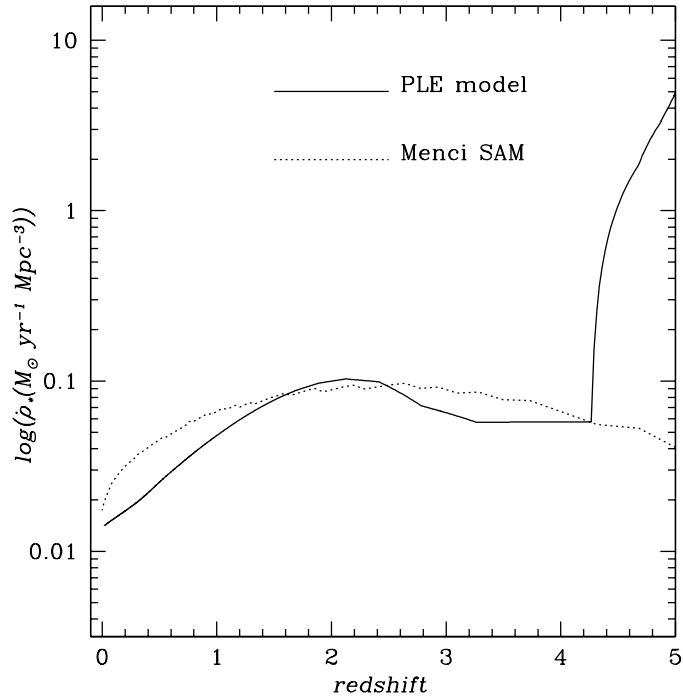,height=10cm,width=10cm}
\caption{ The global SFR density versus redshift as predicted by the PLE model (\emph{solid line}) and by the 
hierarchical SAM (\emph{dotted line}) of Menci et al. 2002. }
\label{SFR}
\end{figure*}

\section{PLE vs SAMs : results}

\subsection{The SFR density}
In Figure \ref{SFR}, we show the evolution of the cosmic SFR density as a function of redshift  
as predicted in the framework of the two scenarios. The two curves have very different shapes: the PLE scenario predicts a 
peak at redshift $z \sim 5$ due to starbursts in spheroids (CM03), followed by a flat behavior between $z\sim 4.2$ 
and $z \sim 3$ due to star formation in spiral galaxies. 
The maximum SF in spirals cause a smaller peak of star formation at $z=2$, 
and these galaxies are the responsible for the decline of the SFR density between  $z=2$ and  $z=0$.\\
The hierarchical SAM model by Menci et al. (2002) produces a curve characterized by a weak increase 
between $z = 5$ and $z \sim 3$, then becomes constant between $z \sim 3$ and $z \sim 2$ and finally starts to 
decrease at $z < 2$ down to $z=0$. Between $z = 2$ and $z = 0$, the SAM model predicts a higher amount of SF than 
the PLE one.

\subsection{The galaxy luminosity density}
\begin{figure*}
\centering
\vspace{0.001cm}
\epsfig{file=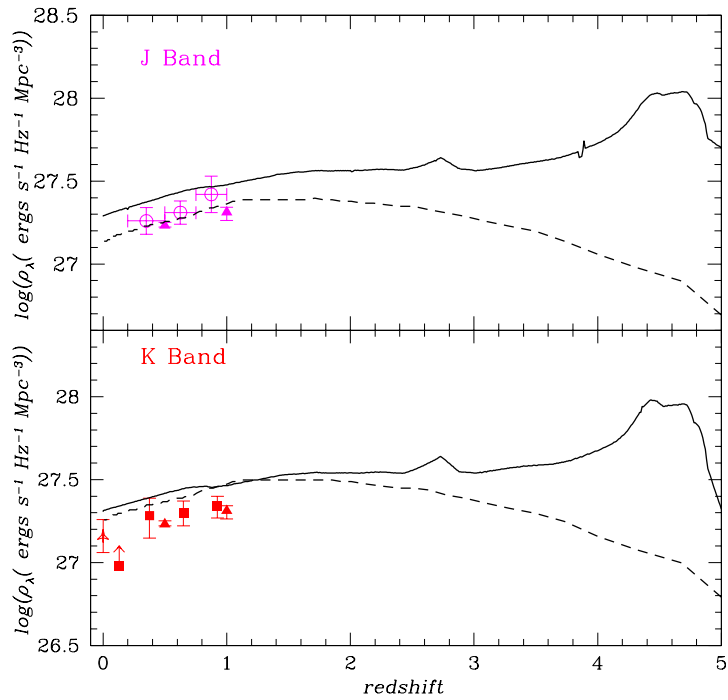,height=10cm,width=10cm}
\caption{Luminosity density evolution in the rest-frame J (upper panel) and K (lower panel) bands as predicted 
by the PLE model by CM03 (\emph{solid curves}) and by the hierarchical SAM of galaxy formation by Menci et 
al. (2002,\emph{dashed curves}), 
and as observed by Lilly et al. (1996, \emph{open circles}), 
Pozzetti et al. (2003, \emph{solid triangles}), 
Gardner et al. (1997, \emph{three-tips stars}), 
Cohen (2002, \emph{solid squares}). }
\label{LD_nodust}
\end{figure*}

In Figure \ref{LD_nodust}, we show the redshift evolution of the luminosity density in the 
 rest-frame K (lower panel) and J (upper panel) bands, 
as predicted by the PLE (solid lines) and by the SAM  
(dashed lines), compared to a set of observational data by various authors.\\ 
The  K band,  centered at  $\lambda =  2.2 \mu$,  is dominated by long
-lived, low mass  stars. The light  emitted in this  band is unaffected  by dust
extinction. 
At $z > 2$,  the two curves have  dramatically different
behaviours: the PLE shows the peak due to ellipticals, whereas the SAM curve has
a broad peak  centered at $z  \sim 2$. On  the other hand,  it is compelling how
similar the curves are at $z<2$. At $z \le 1$ we show the observational data  by
Pozzetti  et al.  (2003) and  Cohen (2002),  in substantial  agreement with  one
another. In this  redshift range, both  curves show broadly a good agreement with  the
observational data.  The PLE  scenario predicts  a slightly  higher LD at $z=0$,
mainly due to  the higher number  of old stars  (hence to redder  galaxy colors)
than the hierarchical picture. From the current set of observational data in the
K band,  it is  practically impossible  to distinguish between the two opposite
galaxy formation scenarios. Rest-frame  Near Infrared deep galaxy  surveys aimed
at detecting faint sources, possibly located at high redshift, could provide  us
with  fundamental hints  to disentangle  between the  PLE and  the hierarchical
scenario.  In fact, if there were an epoch when the bulk of spheroidal  galaxies
is  forming,  the  K-band  LD  would  show  a  peak  centered  at  the  redshift
corresponding to that epoch.  On the other hand, if massive galaxy formation is
distributed throughout an extended  period, no peak in  the K band LD  should be
visible at high redshift.  
These results indicate  that the study of  the evolution of  the  K band
luminosity density at redshift larger than 2 could represent  the
most direct  observational strategy  to establish  the best  scenario of  galaxy
formation. 

Similar  conclusions  can  be  drawn in  the  J  band,  dominated  both by
relatively old stars experiencing the red  giant branch phase and by young  main
-sequence stars and  in very similar fractions (Bruzual 2003).\\ 
The above results, concerning the luminosity density in bands where the contribution 
of long-lived stars is relevant, show that the PLE and SAM models correctly 
predict the total amount of stars formed by  $z\approx 0$, a conclusion 
confirmed by our analysis of the stellar mass density (see below, sect. 4.3). 
The difference between the two scenarios is related to the rate of star formation during the 
cosmic time, which is better probed in the UV and B bands, where the contribution from 
massive, young stars is dominant.  

\begin{figure*}
\centering
\vspace{0.001cm}
\epsfig{file=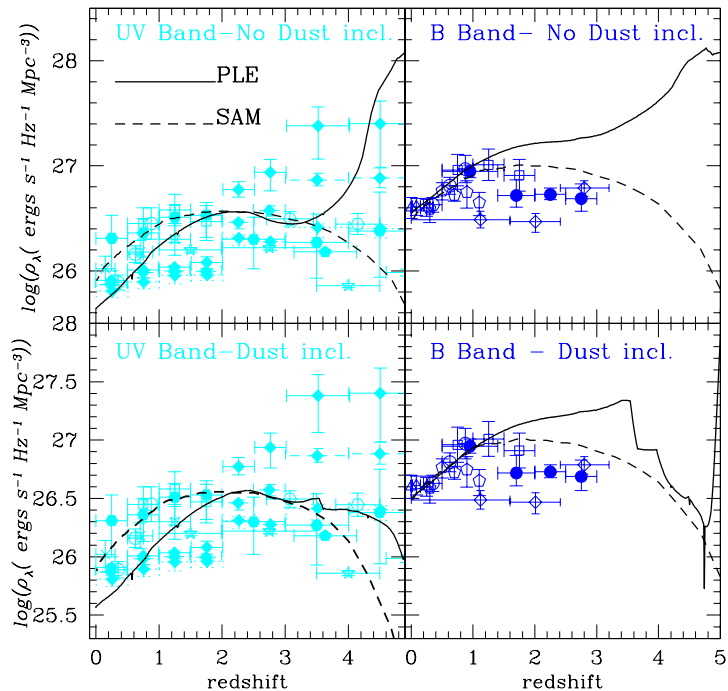,height=10cm,width=10cm}
\caption{Luminosity density evolution in the rest-frame UV and B bands as predicted 
by our PLE models (\emph{solid curves}) and by the hierarchical SAM of galaxy formation by 
Menci et al. (2002) (\emph{dashed curves}) and as observed by varour authors. 
For the UV band, the theoretical curves have been calculated at rest-frame $1400$ ${\rm \AA}$. 
UV band observations: Cowie et al. 
(1999, 2500${\rm \AA}$, \emph{four-tips stars}), Pascarelle et al. (1998, $1500$ ${\rm \AA}$,  
\emph{solid hexagons}), Steidel et al. (1999, $1500$ ${\rm \AA}$, \emph{open hexagons}), Treyer et al. 
(1998, 2000 ${\rm \AA}$, \emph{cross}), Massarotti et al. (2001, 1500 ${\rm \AA}$, \emph{five-tips stars}), 
Giavalisco et al. (2004, 1500 ${\rm \AA}$ \emph{solid pentagons}), Lilly et al. (1996, 2800 ${\rm \AA}$, \emph{open circles}), 
Connolly et al. (1997, 2800 ${\rm \AA}$, \emph{open squares}),  
Lanzetta et al. (2002, 1500 ${\rm \AA}$, \emph{solid diamonds}, plotted for different values of the parameters involved in their measure). 
B band observations: Ellis et al. (1996, $4400$ ${\rm \AA}$, \emph{open triangles}), Dickinson et al. (2003, $4500$ ${\rm \AA}$, 
\emph{solid circles}), Rudnick et al. (2003, $4400$ ${\rm \AA}$, \emph{open diamonds}), Connolly et al. 
(1997, 4400 ${\rm \AA}$, \emph{open squares}), Lilly et al. (1996, 4400 ${\rm \AA}$, \emph{open circles}), 
Wolf et al. (2003, $4560$ ${\rm \AA}$, \emph{open pentagons}). 
In the two upper panels, the theoretical curves are not corrected for dust extinction. In the two lower panels, the curves take into 
account dust extinction corrections. }
\label{LD_dust}
\end{figure*}
\begin{figure*}
\centering
\vspace{0.001cm}
\epsfig{file=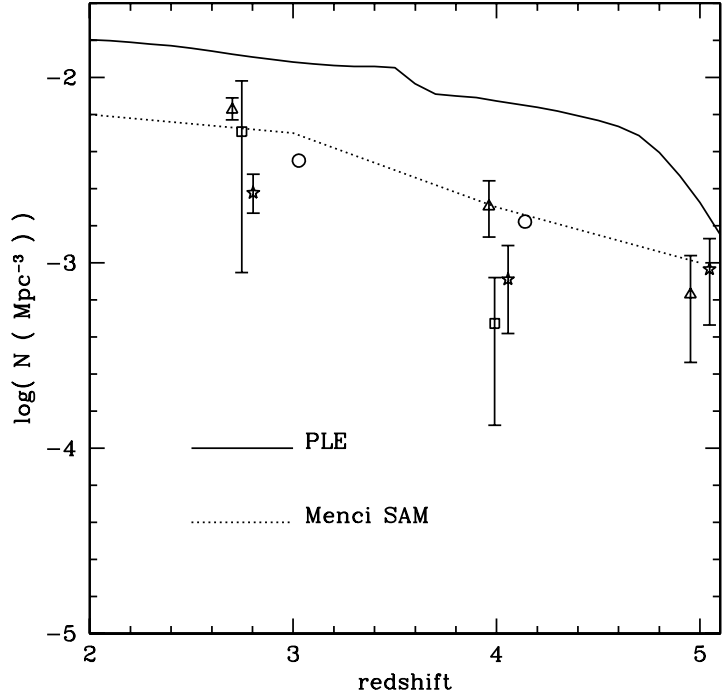,height=10cm,width=10cm}
\caption{Evolution of the total comoving number density of galaxies brighter than 25.5 at rest-frame 
1500${\rm \AA}$ between redshift 2 and 4.9, as predicted by 
by the PLE model (\emph{solid line}) and by the hierarchical SAM of galaxy formation by Menci 
et al. (2002) (\emph{dotted line)}, 
and as observed by Steidel et al.(1999, \emph{open circles}),   
Pozzetti et al. (1998, \emph{open squares}), Lanzetta et al. (1999, \emph{open triangles}), 
Chen et al. (1998, \emph{stars}). This compilation of data has been taken from Somerville et al. 
(2001). For the sake of consistency with the data, in this case we assume a 
$\Lambda$CDM cosmology with $H_{0}=70 km \, s^{-1} \, Mpc^{-1}$. The theoretical predictions have been corrected for dust extinction.}
\label{num}
\end{figure*}
\begin{figure*}
\centering
\vspace{0.001cm}
\epsfig{file=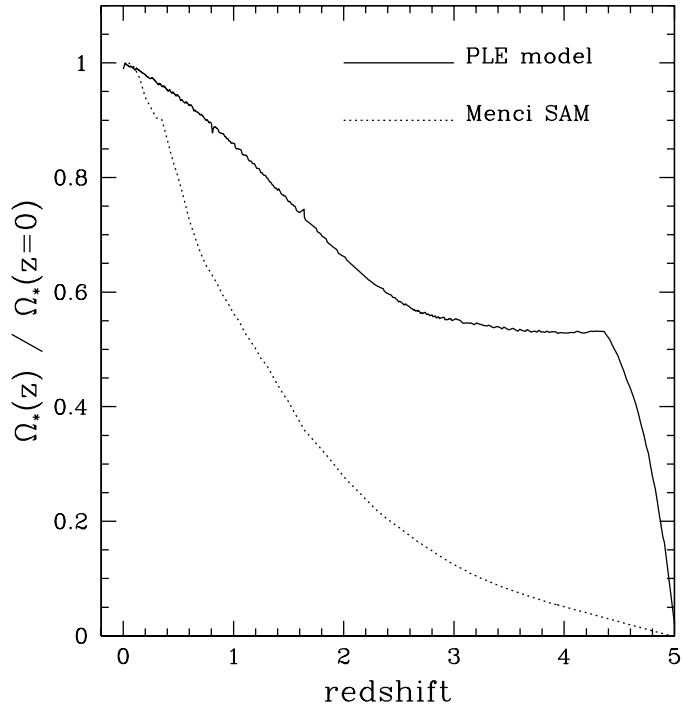,height=10cm,width=10cm}
\caption{Predicted fraction of the total present-day stellar mass as a function of redshift. \emph{Solid line}: PLE model. 
\emph{Dotted line}: SAM by Menci et al. (2002). }
\label{omega_perc}
\end{figure*}
\begin{figure*}
\centering
\vspace{0.001cm}
\epsfig{file=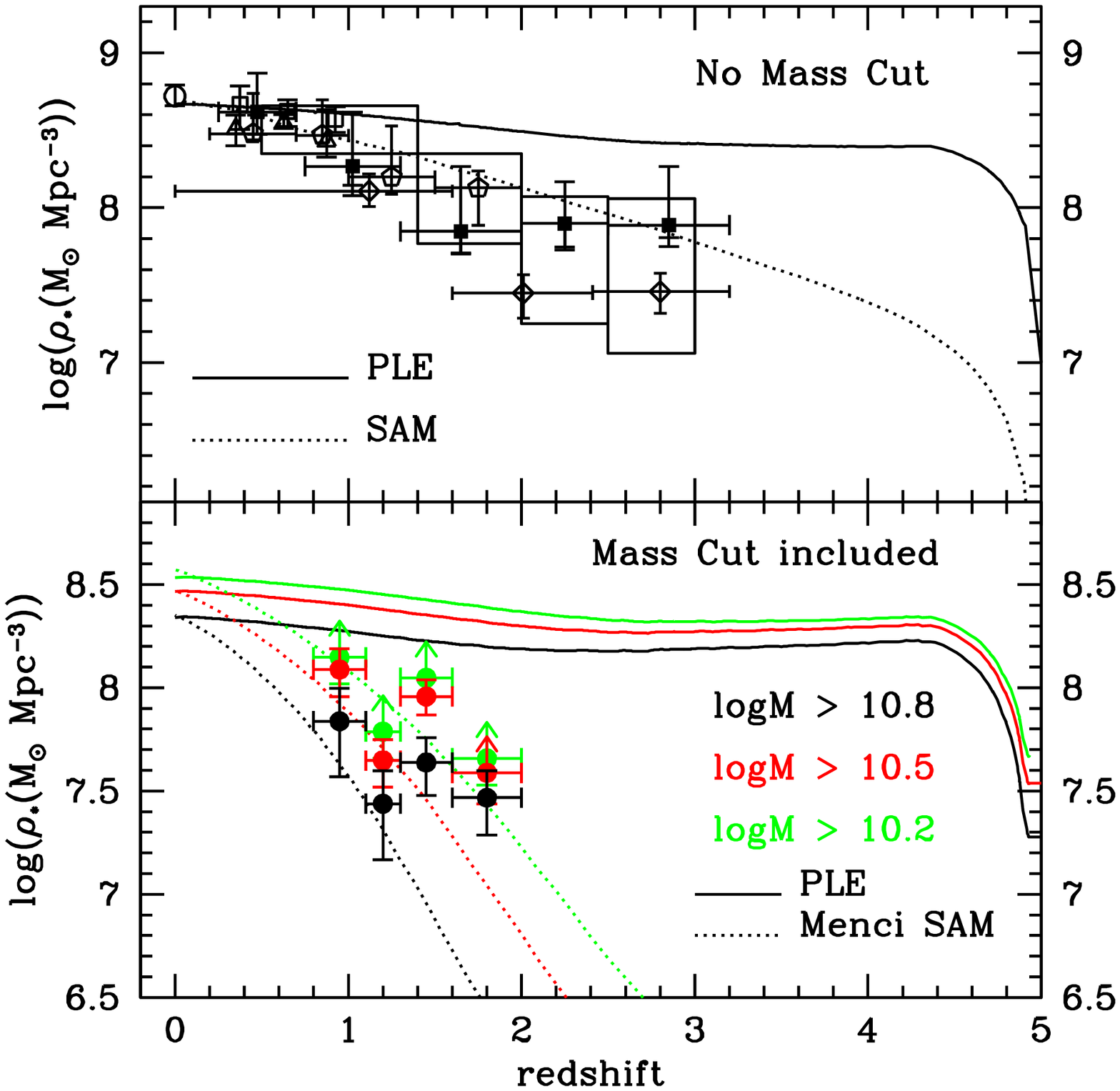,height=10cm,width=10cm}
\caption{\emph{Upper panel}: evolution of the stellar mass density as predicted 
by our PLE models (\emph{solid line}), by the hierarchical SAM of galaxy formation by Menci 
et al. (2002) and as observed by various authors: Dickinson et al. (2003, \emph{square boxes}), 
Cole et al. (2001, \emph{open circle}), 
Brinchmann \& Ellis (2000, \emph{open triangles}), Cohen (2002, \emph{open squares}), Fontana et al. (2003a, \emph{solid squares}), 
Rudnick et al. (2003, \emph{open diamonds}), Fontana et al. (2004, \emph{open pentagons}). 
 No mass cut has been applied to the predicted values. 
\emph{Lower panel}: predicted evolution of the stellar mass density 
according to the PLE (\emph{solid curves}) and SAM (\emph{dotted curves})   
by considering all the stars in galaxies with masses above three mass-cuts, 
namely $M > 10^{10.2} M_{\odot}$ (thick green lines), $M > 10^{10.5} M_{\odot}$ (thick red lines) and $M > 10^{10.8} M_{\odot}$ 
(thick black lines). The predictions are compared with observational values obtained by Glazebrook et al. (2004, \emph{solid circles})   
with the same criteria, i.e. by applying the three same mass cuts to the data sample. 
The values by Glazebrook et al. (2004) corresponding to the three cuts 
are plotted with the same colors as used for the theoretical predictions. }
\label{omega}
\end{figure*}

Figure \ref{LD_dust} shows the evolution of the rest-frame UV and B luminosity density, 
as predicted by the PLE (solid curves) and SAM models (dashed curves).  
In this case, the theoretical LDs have been calculated at $1400{\rm \AA}$ and have been compared  
with data measured at various wavelengths, ranging from $1500$ to $2800$  ${\rm \AA}$ (see caption to Fig.\ref{LD_dust} 
for further details). 
In the two upper panels, the theoretical predictions are not 
corrected for dust extinction, whereas in the two lower panels the curves take into account also corrections for dust-extinction. 
Looking at the upper left panel it is possible to see how, once dust correction is not taken into account,  
in the UV band  the PLE scenario  predicts a strong peak at redshift 5. This peak is due to 
star formation in spheroids, which is absent in the hierachical scenario of Menci et al. (2002).
On the other hand, the SAM curve shows a broad peak, centered at redshift  $\sim 
2.5$. Another   difference concerns  the predicted  evolution at  redshift $<1$,
where  the curve  from the  SAM is  constantly higher  than the  PLE one.  This
reflects the fact that the SAM model predicts a higher amount of star  formation
occurring at $z < 1$ than the PLE curve; this is mainly due to the  contribution
of small-mass galaxies, which  retain a relevant fraction  of their gas down  to
small $z$, while the massive galaxy population, originated from clumps  formed
at high $z$ in  high-density regions, has already consumed most of the  available
cold gas reservoir.\\ The curves calculated in the B band (upper right panel) show a behaviour very
similar to the  UV band, since  both are dominated  by the same  types of stars,
i.e.  the youngest and the most  massive ones. Both bands are sensitive  to dust
extinction, but in a different way:  a comparison with the observations can be
discussed  only after  having corrected  the curves  for dust  obscuration. \\
In the lower left panel of Figure \ref{LD_dust}, the predicted UV luminosity densities have been corrected 
for dust extinction. 
A very important result regarding the UV luminosity density predicted by the 
PLE scenario is that, once dust effects are properly taken into account, the peak at $z \sim 5$ 
due to ellipticals appears considerably reduced, with the PLE curve showing a flat behaviour as the observational data 
by Pascarelle et al. (1998) and Steidel et al. (1999). 
This means that, as suggested by CM03,   
if the bulk of the star formation in the high-redshift universe occurred in sites highly obscured 
by dust, most of it would be invisible for  rest-frame UV surveys (see also Franx et al. 2003). 
Of great interest would be the study of the IR/submm luminosity density, which would be considerably enhanced by the 
re-emission by dust of all the UV absorbed flux, and which is deferred to a forthcoming paper. 
It is also important to 
note that at redshift $>4$, the dust-corrected prediction from the hierarchical model is 
critical: at very high redshift, the unobscured UV luminosity density 
(and hence the amount of star formation) is probably underestimated by the SAM by a factor of  
$3 $ or more, although the scatter in the data is too large to draw firm conclusions.
However, recent independent analysis (Fontana et al. 2003b, Menci et al. 2004) 
have shown that when only the bright galaxy population 
is selected, the paucity of the predicted UV luminosity density compared with observations
is more clearly revealed, confirming that at those $z$ some fundamental process must be 
at work, such as bursts of star formation with a rate higher than that predicted 
by standard SAMs. 
Such a process could be constituted 
by starbursts triggered by interactions of galaxies, as described in Menci et al. (2004) but not included 
in the SAM adopted in this paper.  
These starbursts would speed up the formation of stars in massive galaxies preferentially 
at high $z$ (where the density of galaxies is larger). 
Such starbursts would affect mainly the massive galaxies (due to their larger cross 
section for interactions) and would hence constitute the counterpart of the 
spheroids assumed to form at high-redshift in the PLE model.\\
Of particular interest are the  data by Lanzetta  et al. (2002,  solid
diamonds  in  Figure  \ref{LD_dust}),  who  found  a  monotonically   increasing
behaviour up  to redshift  10. These data take into
account also surface brightness dimming effects, which are likely to be  serious
at high redshift and which have never been considered before by any other group. In
their most extreme case, the observations are as high as the values predicted by
the PLE curve uncorrected for dust. If confirmed by other deep surveys, the data
by Lanzetta et al. (2002) could represent the most direct evidence in favor of a
peak  of star  formation at  high redshift. 
If true, such a peak would be problematic 
to explain for both PLE and hierarchical scenarios. 
However, it also worth stressing that among the three sets of data calculated by Lanzetta et al. (2002) 
the most favored one by the authors is represented by the solid diamonds with dotted error bars, of which the point at redshift $z>4$ 
is in very good 
agreement with the 
PLE predictions but discordant with the SAM predictions. 
\\ Also in the case of the high-redshift
UV LD, the PLE and the hierarchical model used in the present work produce  very
different  predictions,  and  the  observations  clearly  allow us  to discriminate
between the  two.\\ Different  indications seem  to come  from the UV luminosity
density at  $z <  1$. The  prediction from  the SAM  by Menci  et al. (2002) can
nicely  reproduce  the data,  whereas  the PLE  prediction  is lower   than  the
observations.  At  $z=0.2$, where  the  lowest redshift  observations  have been
performed by Treyer et al. (1998) at $\lambda = 2000 {\rm \AA}$, the PLE  models
underestimates  the  data  by  a factor  of $\sim$ 2.2, whereas the data by Lanzetta et al. (2002) 
at $z=0.25$ are underestimated by a factor of $\sim$ 1.2.   
The explanation of this discrepancy is in part related to the fact that in the  morphological classification of the PLE scenario 
we do  not take  into account nearby starburst galaxies, which can contribute up to the $\sim  20\%$ of
the global star formation in the local universe (Brinchmann et al. 2003). This would be enough to account for 
the discrepancy between the PLE predictions and the data by Lanzetta et al. (2002), but not for the data by Treyer et al. (1998). 
\\ 
However, beside the missing contribution by starbursts, also the uncertainty in the B band LF normalization plays an important role. 
The local B band LD adopted here for the PLE model is the one measured by Marzke et al. (1998), 
whose normalization is the lowest among the values provided by the most popular surveys (see Cross et al. 2001) and whose 
uncertainty could reach also factors of $\sim 2$.   
This fact could lead to a slight underestimation of all the LD values predicted by the PLE model.\\ 
The lower right panel of  Figure \ref{LD_dust} shows the  observed evolution of the B 
band luminosity  density  compared  with  the  predictions  corrected  for dust
-extinction. At $z<2$  the PLE and  SAM curves are  overlapping and both  are in
excellent  agreement with  the observations.  At $z  > 2$,  the only  available
measures are the ones by Dickinson et  al. (2003) and by Rudnick et al.  (2003),
none of which are accurately reproduced by any of these scenarios. 
In this case, however, the discrepancy is more critical for the PLE model than for the SAM. 
It is worth to stress that the combination of 
small field, cosmic variance effects,  dust extinction  and 
incompleteness  are a  non-negligible source  of
uncertainty in  the data.  Indeed, some of these effects cause  also an  underabundance of
massive  galaxies as obtained  by  Dickinson  et  al.  (2003)  and a consequent
underestimation of the stellar mass  density with respect to the estimates  by other
authors (Fontana et al. 2003a, see  also section \ref{smdensity}). 
Also in the  B 
band, absorption by dust significantly reduces the peak at $z \sim 4-5$  due to
ellipticals, although to a minor extent than in the case of the UV band. 
In particular, the PLE model predicts a very narrow peak between redshift 5 and 4.8, 
corresponding to a time interval of $\sim 60 Myr $.   
During this interval, the gas in spheroids is experiencing strong metal enrichment, consequently its optical depth is 
progressively rising to its maximum (see eq. 6) and the B band  LD to its minimum.  
The fact that the peak is so narrow is due to the  
assumption that all spheroids start forming stars at the same redshift ($z_{f}=5$) and the star formation is completed after 
$t\simeq 0.3$ Gyr. In a more realistic picture, the first galaxies started  
forming stars before redshift 5 (see Giavalisco et al. 2004) and 
on a finite redshift range, 
so that the very narrow peak would become larger and lower.  
Objects at high redshift which could be associated to a tail in the formation of galactic spheroids 
are the Lyman-break galaxies, which are usually detected at $z \ge 3$ and which show a large range of stellar 
population ages (Papovich et al. 2001, Shapley et al. 2001). In our picture, these galaxies can be associated to  
 forming spheroids (see Matteucci \&   
Pipino 2002), with total stellar masses of the order of the Galactic Bulge. 
Other interesting objects are the submillimeter-bright galaxies, 
detected at z$\sim 2-3$ and characterized by star formation rates of the order of 100- 1000 $M_{\odot}/yr$ 
(Smail et al. 2004). These galaxies have typical space densities of $\sim 10^{-4} Mpc^{-3}$, i.e. comparable 
to $ L_{*}$ ellipticals (Blain et al. 2004). They appear as massive as the largest spheroids observed locally 
and gas-rich (see Neri et al. 2003), and in the PLE picture they can be associated to a tail in the formation of 
massive spheroids. In a $\Lambda$CDM cosmology, the time lag between redshift $2$ and $5$ corresponds to $\sim 2.3$ Gyr. 
This time-spread is consistent with what suggested by Bower et al. (1992), who found that 
in galaxy clusters the redshift range interested by major spheroid formation 
could correspond to an age spread of $\Delta_{form}\sim 2 $ Gyr.  
In the field, Bernardi et al. (1998) found a slightly larger age spread for large spheroids, i.e. $\Delta_{form}\sim 3 $ Gyr.\\ 

Another peak 
is predicted by the  PLE curve at $z  \sim 3.5$, once the interstellar gas 
has completely been ejected by spheroids into the IGM, making the emission by the stars totally visible.\\
Further observations 
in the B band at redshifts of 2-3 and beyond, within the reach of next generation deep galaxy 
surveys, could constitute a stringent test for PLE models. 
If the behaviour shown by present data should be confirmed
by future surveys, this could constitute a strong evidence for 
galaxy density evolution, the process not taken into account in PLE models.

\subsection{The comoving galaxy number density}\label{cndensity}
In Figure \ref{num} we plot the redshift evolution of the number density of bright galaxies. 
Such quantity is obtained by integrating the rest-frame luminosity function at 1500 ${\rm \AA}$,  
considering only the objects brighter than the apparent magnitude limit of $m_{1500} = 25.5$. 
We consider only the redshift range between $z=2$ and $z=5$, i.e. the interval where the predictions  
provided by the PLE and hierarchical scenarios differ most. The observational 
data belong to various authors (see caption to Fig. \ref{num} for further details) and have been all taken 
from Somerville et al. (2001). 
The observations indicate that most of the galaxy number evolution occurs in this redshift range: 
the number of bright galaxies is increasing by a factor of $\sim 6$ between redshift $z=5$ and $z \sim 2.8$. 
The theoretical curves plotted in Figure \ref{num} take into account dust corrections and represent 
the predictions according to the PLE (solid line) and hierarchical (dotted line) scenarios. 
The comparison between the theoretical predictions and the observations considered in this case indicates that 
the PLE scenario is inadequate to describe the number evolution of bright UV galaxies, 
since it systematically overestimates the observed number at all redshifts. We note that the disagreement 
between the PLE curve and the data is maximum at redshift $z\sim 4$, where the discrepancy is of a factor of 
$\sim 5$. On the other hand, the hierarchical scenario described by the SAM allows us to reproduce the observed trend 
with very good accuracy. It is worth noting that the study of the number density of bright UV galaxies represents 
an interesting test for the evolution of star forming galaxies at high redshift but, as well as the UV luminosity density, it does 
not provide any information about the formation of massive spheroids, which most likely occurs in dust-enshrouded 
environments and are thus invisible in the rest-frame UV. Furthermore, if at redshift 3-4 there was already 
a significant number of massive galaxies containing old stars, generating red spectra, 
such population would be certainly missed by UV galaxy 
surveys. A fruitful test for the identification of the number of massive galaxies at high redshift 
is the study of the evolution of the stellar mass density.

\subsection{The evolution of the stellar mass density}\label{smdensity}

Figure \ref{omega_perc} shows the redshift 
evolution of the stellar mass fraction  
as predicted 
by the PLE model (solid line) and by the SAM (dashed line).  
Each curve is normalized to the value for the stellar mass density predicted at the present-day. 
This figure is helpful to understand what percentage of the present-day stellar mass is in place at any given redshift 
according to the predictions of the two scenarios.  
The two curves have a very different behaviour: according to the PLE model, 
nearly half of the stars observable today are already in place at $z\sim 4$, corresponding to 1.63 Gyr after the big 
bang for the cosmology adopted here.  
This is due to the stellar mass produced in spheroids.  
The increase from $z=4$ to $z=0$ is due to quiescent star formation in spirals (CM03). 
At $z=1$, corresponding to an age of the universe of 6.2 Gyr, the PLE model predicts that $85 \%$ of the present stellar 
mass is already in place.\\ 
According to the hierarchical SAM, the buildup of the stellar mass occurs progressively, with  
half of the total stellar mass in place at $z \sim 1.2$, i.e. 5.42 Gyr after the big bang. 
By $z \sim 1$, the SAM predicts that nearly $60 \%$ of the total present stellar 
mass is present.\\ 
Figure  \ref{omega}  shows a  comparison  between the  stellar  mass density  as
observed by various authors and as predicted by PLE models and by the SAM.  This
comparison demonstrates that, owing to  the extreme differences between the  PLE
and SAM  predictions, the observation of  the stellar mass  density constitutes
another  very  helpful strategy  to   distinguish  between  the  PLE  and   the
hierarchical scenario. \\ 
In the upper panel of figure \ref{omega}, we show the evolution of the stellar mass density by considering 
galaxies of all masses, namely no mass cut has been applied to the predicted values. The theoretical predictions are 
compared with observational estimates by various authors (for further details, see caption of Fig. \ref{omega})
In general, the main sources of uncertainties in the data are 
dust extinction and cosmic variance effects due to  small
field. The data by Fontana et al. (2003a) are taken from a large volume and  are
corrected for dust extinction. However, as emphasized by the authors,  they may
still suffer for incompleteness on the bright end of the mass function. 
To estimate to what extent these effects could alter the real values is difficult: for instance,  
the amounts of dust can vary considerably from one galaxy to another. 
Also the cosmic variance effects are in principle difficult to evaluate. 
It is worth noting that all these effects conspire to lower the estimates of the stellar mass at 
redshifts larger than 1: for these reasons, it is safe to consider the data as lower limits. 
The PLE and SAM curves are both in reasonable agreement with the data within redshift $z < 1.5$. 
At redshifts higher than $1.5$, if we consider the predicted total stellar mass the PLE model presents  
a noticeable discrepancy with the observations: if we consider the central values estimated by Fontana et al. (2003), 
the discrepancies between observations and PLE predictions are by factors of $3 - 6$. 
On the other hand, on average, the SAM predictions seem to show a good agreement with the observed values.\\
In the lower panel of figure \ref{omega}, we show the predicted evolution of the stellar mass density 
according to the PLE (solid curves) and SAM (dotted curves) and  
by considering all the stars in galaxies with masses above three mass-cuts, 
namely $M > 10^{10.2} M_{\odot}$ (thick green lines), $M > 10^{10.5} M_{\odot}$ (thick red lines) and $M > 10^{10.8} M_{\odot}$ 
(thick black lines). Such predictions are compared with observational values obtained by Glazebrook et al. (2004) 
with the same criteria, i.e. by applying the same three mass cuts to the data sample. 
The values by Glazebrook et al. (2004), corresponding to the three cuts,  
are plotted with the same color as used for the theoretical predictions. 
The adoption of the mass cuts is very helpful in establishing a 
full correspondence between observations and theoretical predictions, and to have a very clear picture of 
the number of massive galaxies that the PLE and hierarchical scenarios predict at any redshift, respectively. 
If we compare the PLE predictions with the data calculated with the three cuts, we notice that the agreement between 
data and predictions does not improve and that the PLE model in general 
tends to overestimate the stellar mass density in massive galaxies, in particular at redshifts $z > 1$.\\ 
If we compare the SAM predictions to the data, we notice that the hierarchical picture can reproduce the observed data 
with the three cuts up to redshift $z \sim 1.2$, whereas at higher redshift 
it tends to underestimate the observations. The disagreement is particularly strong for the highest mass cut 
($M > 10^{10.8} M_{\odot}$). 
This shows that at redshifts $ z\ge 1$,
according  to the  SAM predictions,  the bulk  of the  stellar mass  resides in 
objects with masses $M<10^{10.2} M_{\odot}$.  These small objects would be
too faint to be visible by  any current high-redshift survey. 
Also in  this case,  this problem  is alleviated by considering the 
effect of interaction-driven starburst in massive  galaxies at high-redshift,  
(see Menci et al. 2004), which would increase 
the fraction of stellar mass already in place at $z= 2$ to a value 
around 0.3 of the present mass density.\\
It is very interesting to see how, by means of $\Lambda$CDM  cosmological numerical simulations, 
Nagamine et al. (2004) find a strong discrepancy between the predicted and observed 
amount of stellar mass at redshift $z > 1.5$. 
Their simulations indicate an excess of stellar mass with respect to observational estimates at high redshift, 
in analogy with the result of the PLE model considered in this work. 
This is another indication suggesting that the global star formation of the universe may have proceeded in the past 
at levels somewhat higher than predicted by semi-analytical models, and it confirms that effects such as dust obscuration 
and cosmic variance may still seriously prevent us from having a clear picture of galaxy evolution at redshifts $z >1$.\\
Recently, the Great Observatories Origins Deep Survey has provided evidence for a population of galaxies showing distorted 
morphologies and with ongoing merger activity located at $z \ge 1.5$ (Somerville et al. 2004). 
The number density of such bright objects is underestimated by current hierarchical SAMs and overestimated by PLE models. 
To assess the role of such galaxies in the stellar and metal budget would be of primary interest in order 
to have further crucial hints on the evolution of galaxies at redshifts larger than 1. 

\section{Conclusions}
In this paper we have studied the evolution of the cosmic star formation, the galaxy luminosity density 
and the stellar mass density by means of two opposite galaxy evolution pictures: 
the pure-luminosity evolution model developed by CM03 and the 
semi-analytical model of hierarchical galaxy formation by Menci et al. (2002). 
The former  predicts a peak  at redshift $z=5$, due to intense  star formation
in ellipticals, followed by a  phase of quiescent and continuous  star formation
occurring in  spiral galaxies.  The SAM  predicts a  smoother behavior, 
following the gradual build up of galactic DM halos through repeated merging events. 
The aim was to derive constraints on the relative importance of different physical 
processes - like the dependence on morphology of the star formation history, 
the density evolution of the galaxy population, the impulsive star bursts - 
in determining the observed properties of the galaxies. 

We have shown that  the evolution of the  cosmic star formation rate  density in
the two models behaves  quite differently. 
However, the integral of the cosmic star formation rate 
at redshift $z\le 1$, probed by the stellar mass density evolution in this redshift range,  
are in good agreement. This ensures that the total amount of stars formed 
along  the star  formation histories  are similar  (and in  agreement with  the
observations). To  probe the rate of star  formation at different 
cosmic epochs we investigated the luminosity density in the UV and B bands, where
the emission is dominated by  young, short-living massive stars. The  comparison
with the {\it available} data shows that: 

1) At redshift $z> 4$, the SAM tends to underestimate the observed UV luminosity
density which, as several current surveys indicate, is a non-decreasing function of 
$z$. On the other hand, the PLE predictions can fairly account for such observed trend. 
If future surveys will confirm such behaviour, this could indicate that some fundamental processes  should 
be inserted into SAM to boost the star formation at high redshifts. 
An example of such a process could be the interaction-driven starbursts suggested by Menci et al. (2004). 

2) In the B band the PLE model tends to overestimate the observed luminosity
density  at $z> 2.5$ by a factor increasing with $z$. This is the 
consequence of placing  a rapid formation of all the elliptical galaxies at
$z\approx 5$. While  dust extinction and incompleteness severely 
affect the comparison with present data,  if future observations will
not indicate a substantial growth of the B-band luminosity density for 
$z\ge 2.5-3$, this would point toward a galaxy density evolution, the 
main process not included in PLE models. 

3) At low redshift ($z<1$), the local UV luminosity density predicted by the SAM 
is about 2 times larger than those arising from PLE models.  
This is because in hierarchical scenarios at low redshift the small mass galaxies 
still retain a significant fraction of their cold gas reservoirs, while the massive 
ones have already exhausted most of their fuel at high redshift, since the latter 
are formed from clumps originated in biased high density regions of the cosmic 
density field. In hierarchical models, at low $z$ the contribution of low-mass galaxies sustains the 
global star formation rate above the value 
obtained in the continuous, passive evolution PLE models. The above discussion
shows that, 
while the local J and K observations will hardly contribute to discriminate between the 
two scenarios, accurate measurements of the local UV luminosity density 
would be effective in constraining the models. 

4) The observed evolution of the comoving number density of bright galaxies at redshift $z\ge 3$ is well reproduced 
by the hierarchical SAM, whereas, for the set of data considered here, the PLE overestimates the observed densities 
by factors between 2 and 5.

5) The stellar  mass density constitutes a complementary probe for the 
PLE and hierarchical  scenarios. In general, both  the PLE  and
hierarchical predictions allow us to reproduce the observed stellar mass density
evolution  up to  $z=1$. At $z>1$, the predicted stellar mass densities diverge, 
with the PLE predictions remaining almost constant up to redshift $z\approx 4$ 
and the SAM predictions continuously dropping with increasing $z$.  
Without any mass-cut on the theoretical predictions, the PLE model overestimates the data by factors of 3-6.  
If we calculate the stellar mass density evolution and apply the three mass cuts, as performed 
by Glazebrook et al. (2004), in general 
the discrepancies between the PLE model and the observations at $ z > 1 $ do not reduce. 
On the other hand, the hierarchical picture  
underestimates the observations for all the three values of the mass-cuts at redshifts $z > 1.2$. 
This is related to the fact that,  
at redshifts $ z\ge 1$, 
according  to the  SAM predictions  the bulk  of the  stellar mass  resides in 
objects with masses $M<10^{10.2} M_{\odot}$.  These small objects would be
too faint to be visible by  any current high-redshift survey. 
Also  in  this  case, the
discrepancy  between  the  hierarchical  model  and  observations  is  partially
alleviated by introducing  a population of  high-redshift starbursts in  massive
galaxies (Menci et al. 2004), which would bring the mass density at $z=2$ to 
values around 1/3 of the local value, in much better agreement with the data but still
well below the PLE predictions. 
Thus, in principle, more precise observations 
of the stellar mass density at $z>2$ will be able to discriminate between
the PLE models and the SAM including starbursts at high $z$.  
On the other hand, some indications against hierarchical formation of elliptical galaxies 
is provided by chemical constraints, in particular the increase of the [Mg/Fe] ratio 
with galaxy luminosity (Pipino \& Matteucci 2004, Thomas 1999 ). This fact indicates that the most massive ellipticals stopped forming stars 
before the less massive ones. All of these facts together will have to be taken into account 
eventually before drawing firm conclusions. \\ 
As forthcoming  work, to 
investigate star and massive  galaxy  formation  at high  redshift we  will use
other diagnostics, such as IR and submm emission.

\section*{Acknowledgments}
We thank an anonymous referee for several enlightening suggestions which improved the 
quality of this work.  
We wish to thank Daniela Calzetti for many useful suggestions on the treatment of dust extinction. 
We thank Cristina Chiappini and Paolo Tozzi for careful readings of the manuscript and for several useful comments.  
F. C. and F. M. also acknowledge funds from MIUR, COFIN 2003, prot. N. 2003028039.

\label{lastpage}


\begin{thebibliography}{99}
\bibitem[]{} Arimoto, N., Yoshii, Y., 1987, A\&A, 173, 23
\bibitem[]{} Baugh, C. M., Cole, S., Frenk, C. S., Lacey, C. G., 1998, ApJ, 498, 504
\bibitem[]{} Barger, A., et al., 1999, AJ, 117, 102
\bibitem[]{} Bernardi, M., et al., 1998, ApJ, 508, L43
\bibitem[]{} Binney, J., Merrifield, M., 1998, ``Galactic Astronomy'', Princeton University Press (Princeton series in astrophysics)
\bibitem[]{} Blain, A. W., Chapman, S. S., Smail, I., Ivison, R., 2004, ApJ, in press, astro-ph/0405035 
\bibitem[]{} Bond, J.R., Cole, S., Efstathiou, G., Kaiser, N., 1991, ApJ, 379, 440
\bibitem[]{} Bower, R. G., 1991, MNRAS, 248, 332
\bibitem[]{} Bower, R. G., Lucey, J. R., Ellis, R. S., 1992, MNRAS, 254, 613
\bibitem[]{} Bradamante, F., Matteucci, F., D'Ercole, A., 1998, A\&A, 337, 338
\bibitem[]{} Brinchmann, J., Ellis, R. S., 2000, ApJ, 536, 77
\bibitem[]{} Brinchmann, J., Charlot, S., White, S. D. M., Tremonti, C., Kauffmann, G., Heckman, T., Brinkmann, J., 2003, MNRAS, submitted,astro-ph/0311060
\bibitem[]{} Bruzual,  A. G., 2003, in ``Galaxies at high redshift'',  XI Canary Islands Winter School of Astrophysics, edited by I. P\'erez Fournon et al., Cambridge University Press, p. 185, astro-ph/0011094
\bibitem[]{} Bruzual,  A. G., Charlot, S., 2003, MNRAS, 344, 1000
\bibitem[]{} Bundy, K., Fukugita, M., Ellis, R. S., Kodama, T., Conselice, C. J., 2004, ApJ, 601, L123
\bibitem[]{} Calura, F., Matteucci, F., 2003, ApJ, 596, 734 (CM03)
\bibitem[]{} Calura, F., Matteucci, F., 2004, MNRAS, in press, astro-ph/0401462
\bibitem[]{} Calura, F., 2004, PhD thesis, Trieste University
\bibitem[]{} Calzetti, D., Kinney, A. L., Storchi-Bergmann, T., 1994, ApJ, 429, 582
\bibitem[]{} Calzetti, D., 1997, in ``The Ultraviolet Universe at Low and High Redshift : Probing the Progress of Galaxy Evolution'', William H. Waller et al. eds., AIP Conference Proceedings, 408, 403
\bibitem[]{} Calzetti, D., 2001, PASP, 113, 1449 
\bibitem[]{} Chen, H.-S., Fernandez-Soto, A., Lanzetta, K. M., Pascarelle, S. M., Puetter, R, C., Yahata, N., Yahil, A., 1998, astro-ph/9812339
\bibitem[]{} Chiappini, C., Matteucci, F., Gratton, R. 1997, ApJ, 477, 765 
\bibitem[]{} Chiappini, C., Matteucci, F., Romano, D., 2001, ApJ, 554, 1044
\bibitem[]{} Cohen, J. G., 2002, ApJ, 567, 672
\bibitem[]{} Conselice, C. J., Bershady, M. A., Dickinson, M., Papovich, C., 2003, AJ, 126, 1183
\bibitem[]{} Connolly, A. J., Szalay, A. S., Dickinson, M. E., SubbaRao, M. U., Brunner, R. J., 1997, ApJ, 486, L11
\bibitem[]{} Cole, S., Lacey, C. G., Baugh, C. M., Frenk, C. S., 2000, MNRAS, 319, 168
\bibitem[]{} Cole, S., et al., 2001, MNRAS, 326, 255
\bibitem[]{} Cowie, L., Songaila, A., Barger, A., 1999, AJ, 118, 603
\bibitem[]{} Cross, N., et al., 2001, MNRAS, 324, 825
\bibitem[]{} Dickinson, M., Papovich, C., Ferguson, H. C., Baudav\'ari, T., 2003, ApJ, 587, 25
\bibitem[]{} Djorgovski, S., Davis, M., 1987, ApJ, 313, 59
\bibitem[]{} Ellis, R. S., Colless, M., Broadhurst, T., Heyl, J., Glazebrook, K., 1996, MNRAS, 280, 235
\bibitem[]{} Fontana, A.; Donnarumma, I.; Vanzella, E.; Giallongo, E.; Menci, N.; Nonino, M.; Saracco, P.; Cristiani, S.; D'Odorico, S.; Poli, F, 2003a, ApJ, 594, L9
\bibitem[]{} Fontana, A.; Poli, F.; Menci, N.; Nonino, M.; Giallongo, E.; Cristiani, S.; D'Odorico, S., 2003b, 
ApJ, 587, 544F
\bibitem[]{} Fontana, A., et al., 2004, A\&A, in press, astro-ph/0405055
\bibitem[]{} Franx, M., et al., 2003, ApJ, 587, L79
\bibitem[]{} Franceschini, A., Silva, L., Fasano, G., Granato, L., Bressan, A., Arnouts, S., Danese, L., 1998, ApJ, 506, 600
\bibitem[]{} Gardner, J. P., Sharples, R. M., Frenk, C. S., Carrasco, B. e., ApJ, 1997, 480, L99
\bibitem[]{} Giavalisco, M., et al., 2004, ApJ, 600, L103
\bibitem[]{} Glazebrook, K., et al., 2004, ApJ, submitted, astro-ph/0401037
\bibitem[]{} Im, M., Griffiths, R. E., Ratnatuga, K. U., Sarajedini, V. L., 1996, ApJ, L79  
\bibitem[]{} Im, M., et al., 2002, ApJ, 571, 136 
\bibitem[]{} Kauffmann, G., White, S. D. M., Guiderdoni, B., 1993, MNRAS, 264, 201
\bibitem[]{} Kauffmann, G., Charlot, S., White, S., 1996, MNRAS, 283, 117
\bibitem[]{} Kennicutt, R. C., 1998, ApJ, 498, 541
\bibitem[]{} Kochanek, C. S., Falco, E. E., Impey, C. D., Leh\'ar, J., McLeod, B. A., Rix, H.-W., Keeton, C. R., Mu\~noz, J. A., Peng, C. Y., 2000, ApJ, 543, 131
\bibitem[]{} Kodama, T., Bower, R. G., Bell, E. F., 1999, MNRAS, 306, 561
\bibitem[]{} Lacey, C., \& Cole, S., 1993, MNRAS, 262, 627
\bibitem[]{} Lanzetta, K. M., Chen, H-S., Fernandez-Soto, A., Pascarelle, S., Puetter, R., Yahata, N., Yahil, A., 1999, in "Photometric Redshifts and High Redshift Galaxies", eds. R. Weymann, L. Storrie-Lombardi, M. Sawicki \& R. Brunner, astro-ph/9907281
\bibitem[]{} Larson, R. B., 1974, MNRAS, 166, 585
\bibitem[]{} Lilly, S. J., LeF\'evre, O., Hammer, F., Crampton, D., 1996, ApJ, 460, L1
\bibitem[]{} Lilly, S., et al., 1998, ApJ, 500, 75
\bibitem[]{} Matteucci, F., Tornamb\'{e}, A., 1987, A\&A, 185, 51
\bibitem[]{} Matteucci, F., Fran\c cois, P., 1989, MNRAS, 239, 885
\bibitem[]{} Matteucci, F., 1992, ApJ, 397, 32
\bibitem[]{} Matteucci, F., 1994, A\&A, 288, 57
\bibitem[]{} Matteucci, F., Pipino, A., 2002, ApJ, 569, L69 
\bibitem[]{} Menanteau, F., Ellis, R. S., Abraham, R. G., Barger, A. J., Cowie, L. L., 1999, MNRAS, 309, 208
\bibitem[]{} Menci, N., Cavaliere, A., Fontana, A., Giallongo, E., Poli, F., 2002, ApJ, 575, 18
\bibitem[]{} Menci, N., Cavaliere, A., Fontana, A., Giallongo, E., Poli, F., Vittorini, V., 2004, ApJ, in press
\bibitem[]{} Marzke, R. O., Nicolaci Da Costa, L., Pelligrini, P., Willmer, C. N. A., Geller, M., 1998, ApJ, 503, 617 
\bibitem[]{} Massarotti, M., Iovino, A., Buzzoni, A., 2001, ApJ, 559, L105
\bibitem[]{} Nagamine, K., Cen, R., Hernquist ,L., Ostriker, J. P., Springel, V., 2004, ApJ, in press, astro-ph/0311294
\bibitem[]{} Neri, R., Genzel, R., Ivison, R. J., Bertoldi, F., Blain, A. W., Chapman, S. C., Cox, P., Greve, T. R., 
Omont, A., Frayer, D. T., 2003, ApJ, 597, L113
\bibitem[]{} Nomoto, K., Hashimoto, M., Tsujimoto, T., Thielemann, F. K., Kishimoto, N., Kubo, Y., Nakasato, N., 1997a, Nucl. Phys. A, 616, 79   
\bibitem[]{} Nomoto, K., Iwamoto, K., Nakasato, N. T., et al., 1997b, Nucl. Phys. A, 621, 467
\bibitem[]{} Patton, D. R., Pritchet, C. J., Yee, H. K. C., Ellingson, E., Carlberg, R. G., 1997, ApJ, 475, 29
\bibitem[]{} Lanzetta, K. M., Yahata, N., Pascarelle, S., Chen, H., Fern\'andez-Soto, A., 2002, ApJ, 570, 492
\bibitem[]{} Le F\'evre, O., et al., 2000, MNRAS, 311, 565
\bibitem[]{} Papovich, C., Dickinson, M., Ferguson, H. C., 2001, ApJ, 559, 620 
\bibitem[]{} Pascarelle, S. M., Lanzetta, K. M., Fern\'andez-Soto, A., ApJ, 1998, 508, L1
\bibitem[]{} Peebles, P. J. E., 2003, in ``A new era in cosmology'', ASP conference series, eds. N. Metcalfe \& T. Shanks, in press, astro-ph/0201015 
\bibitem[]{} Pipino, A., Matteucci, F., 2004, MNRAS, in press, astro-ph/0310251
\bibitem[]{} Poggianti, B.M., 1997, A\&ASS, 122, 399
\bibitem[]{} Poli, F., Giallongo, E., Menci, N., D'Odorico, S., Fontana, A., 1999, ApJ, 527, 662
\bibitem[]{} Pozzetti, L., Madau, P., Zamorani, G., Ferguson, H. C., Bruzual, G. A., 1998, MNRAS, 298, 1133
\bibitem[]{} Pozzetti, L., et al., 2003, A\&A, 402, 837
\bibitem[]{} Press, W.H., Schechter, P., 1974, ApJ, 187, 425 
\bibitem[]{} Renzini, A., Ciotti, L., 1993, ApJ, 416, L49
\bibitem[]{} Renzini, A., 1999, in "When and How do Bulges Form and Evolve?", ed. by C.M. Carollo, H.C. Ferguson \& R.F.G. Wyse (Cambridge University Press), astro-ph/9902108 
\bibitem[]{} Rudnick, G., et al., 2003, ApJ, 599, 847
\bibitem[]{} Rusin, D.,  et al., 2003, ApJ, 587, 143
\bibitem[]{} Salpeter, E. E., 1955, ApJ, 121, 161
\bibitem[]{} Sandage, A., 1986, A\&A, 161, 89
\bibitem[]{} Scalo, J. M., 1986, FCPh, 11, 1
\bibitem[]{} Schade, D., et al., 1999, ApJ, 525, 31
\bibitem[]{} Schechter, P., 1976, ApJ, 203, 297
\bibitem[]{} Seaton, M. J., 1979, MNRAS, 187, 73
\bibitem[]{} Shapley, A. E., Steidel, C. C., Adelberger, K. L., Dickinson, M., Giavalisco, M., Pettini, M., 2001, ApJ, 562, 95
\bibitem[]{} Smail, I., Chapman, S., Blain, A., Ivison ,R., 2004, in "Maps of the Cosmos", IAU Symposium 216, ASP Conference Series, eds. M. Colless \& L. Staveley-Smith, Sydney, July 2003, 11 pages, astro-ph/0311285
\bibitem[]{} Somerville, R. S., Primack, J. R., Faber, S. M., 2001, MNRAS, 320, 504
\bibitem[]{} Somerville, R. S., et al., 2004, ApJ, 600, L135
\bibitem[]{} Springel, V., Hernquist, L., 2003, MNRAS, 339, 312
\bibitem[]{} Steidel, C. C., Giavalisco, M., Pettini, M., Dickinson, M., Adelberger, K. L., 1996, ApJ, 462, L17
\bibitem[]{} Thomas, D., 1999, MNRAS, 306, 655 
\bibitem[]{} Totani, T., Yoshii, Y., 2000, ApJ, 540, 81 
\bibitem[]{} Treyer, M. A., Ellis, R. S., Milliard, B., Donas, J., Bridges, T. J., 1998, MNRAS, 300, 303
\bibitem[]{} van Albada, T. S., 1982, MNRAS, 201, 939
\bibitem[]{} van den Hoeck, L. B. \& Groenwegen, M. A. T., 1997, A\&AS, 123, 305
\bibitem[]{} van Dokkum, P. G.; Franx, M.; Fabricant, D.; Illingworth, G. D.; Kelson, D. D., 2000, ApJ, 541, 95
\bibitem[]{} van Dokkum, P. G., Franx, M., Kelson, D. D., Illingworth, G. D., 2001, ApJ, 553, L39
\bibitem[]{} van Dokkum, P. G., Ellis, R. S., 2003, ApJL, in press, astro-ph/0306474
\bibitem[]{} White, S. D. M., Rees ,M. J., 1978, MNRAS, 183, 341
\bibitem[]{} Wolf, C.,  Meisenheimer, K., Rix, H.-W., Borch, A., Dye, S., Kleinheinrich, M., 2003, A\&A, 401, 73
\bibitem[]{} Zepf, S. E., 1997, Nat., 390, 377
\end{thebibliography}
\end{document}